# Design of Carrier-Depletion-based Modulators on Commercially Available SOI Substrates of Varying Device-Layer Thickness

Anjali A R[1,2], Syamsundar De[1], and Pranabendu Ganguly[1,3]

[1]*Indian Institute of Technology Kharagpur, Advanced Technology Development Centre, West Bengal-721302, India*

[2]*anjaliar@kgpian.iitkgp.ac.in*, [3]*pran@atdc.iitkgp.ac.in*

**Abstract:** We compare modulator designs on thick-film and thin-film silicon-on-insulator (SOI) substrates based on the carrier-depletion effect. This effect exhibits a lower junction capacitance as compared to its injection counterparts, leading to a higher electro-optic bandwidth, although the effective refractive index change is low. In this work, commercially available standard SOI substrates with device layer thicknesses of 5 µm, 3 µm, and 0.22 µm are chosen for designing carrier-depletion-based phase shifters, which are then utilized for the design of Mach-Zehnder modulators (MZMs). These MZMs are tested for various on-off-keying (OOK) modulation bit rates employing non-return-to-zero (NRZ) pseudo-random binary signal (PRBS). For the comparison of given designs, we find that the thick-film MZMs, which can provide higher fabrication tolerance and high power handling capabilities, support maximum modulation speeds in the order of a few Gbps, whereas it can go up to 20 Gbps for thin-film MZMs.

## 1. Introduction

The maturity of the silicon microelectronics industry led to the use of silicon platforms for photonics applications. This trend is further enhanced after Soref et al. [1] reported the electro-optic effects in silicon, leading to active integrated photonic devices. Silicon photonics is predominantly realized on the silicon-on-insulator (SOI) platform, which offers high index contrast and CMOS compatibility. Early implementations employed thick-film SOI substrates, with device layers ranging from 3 to 5 µm, suitable for butt-coupling and early optical interconnects [2, 3]. However, with increasing demands for compactness, higher bandwidth, and energy efficiency, the field transitioned toward thin-film SOI platforms, notably the 220 nm device layer, which enables strong mode confinement and efficient phase modulation through compact structures [4, 5]. The development of electro-optic modulators (EOMs) on SOI platforms quickly became popular because modulators are an integral part of optical interconnects for high-speed data transfer. Both carrier injection and carrier depletion mechanisms have been extensively explored in SOI-based modulators [6, 7, 8, 9, 10]. Carrier injection, which relies on forward-biased p-n or p-i-n junctions, offers a large refractive index change, but suffers from limited bandwidth due to minority carrier lifetimes. In contrast, carrier depletion, which employs reverse-biased junctions, provides faster modulation speeds at the expense of a smaller index change. Despite this trade-off, carrier-depletion-based modulators have become the preferred architecture for high-speed applications, especially in Mach-Zehnder interferometer (MZI)-based designs due to their linearity and scalability [11, 12]. All-silicon phase and intensity modulators on both thick and thin SOI substrates have been extensively studied. For example, in thick-film SOI, early reports showed modulators employing both carrier depletion [13] and carrier injection [14] with moderate bandwidths of only a few MHz. However, experimental validation and in-depth studies of thick-film SOI modulators remain relatively scarce. On the other hand, 220-nm SOI platform employing carrier depletion achieved state-of-the-art modulators having $V_{\pi}L$ <1 V·cm, electro-optic bandwidths >30 GHz, and extinction ratios (ERs) >20 dB [15, 16]. In particular, for all silicon MZMs, reported results have demonstrated bandwidths above 100 GHz with on-off keying (OOK) modulation [17], extinction ratio values above 40 dB [8, 18], data rates reaching up to 224 Gbps [19], and insertion losses less than 5 dB [20]. Although there have been fewer studies on carrier-depletion modulators using thick-film SOI platforms, these platforms are valuable for applications that need low-cost solutions and lower-density integration. Currently, there is a lack of comprehensive comparisons between modulators built on thick-film and thin-film SOI substrates, which leaves a knowledge gap. Our work focuses on designing depletion-based Mach Zehnder Modulators (MZMs) on thick-film SOI to address this gap by examining the performance trade-offs and practical benefits of using thick-film technology. This is important because thick-film SOI enables designs that are better suited for high-power and relaxed bandwidth conditions, while also broadening the scope of modulator research beyond the predominantly thin-film SOI field. By eliminating any form of hybrid integration, we focus our analysis on the performance differences arising from the thickness of the device layer and the geometry of the waveguide. We analyze and benchmark the achievable modulation efficiency, extinction ratio, and bandwidth across the two platforms, aiming to understand the trade-offs and optimization strategies for intensity modulators using the carrier depletion effect. The paper is organized as follows: Section 2 describes the device design and the operating principle, followed by the numerical simulation-based performance analysis of the devices discussed in Section 3.

## 2. Device Modeling

### 2.1 Rib waveguide design

For designing phase shifters on both thick- and thin-film silicon-on-insulator (SOI) substrates, the initial step is to optimize the waveguide dimensions to support single-mode propagation. In this work, the Lumerical MODE solutions tool was used for this purpose. At the targeted wavelength of 1550 nm, the refractive index for silicon (Si) is taken as 3.475, and the index for silicon dioxide (SiO2) is 1.4447. The Si rib waveguide is embedded in an outer cladding of SiO2. The final dimensions are listed in Table 1, and the substrates are designated as I, II, and III throughout the article. It is important to note that the designed thick film-based rib waveguides support both fundamental transverse electric (TE) and transverse magnetic (TM) modes, due to their polarization-dependent nature, whereas in the case of a thin film substrate, it is possible to achieve the single-mode condition with only the fundamental TE mode. Henceforth, we consider only the TE mode for comparative analysis of the different substrates. The effective refractive indices of the fundamental TE mode are 3.4706, 3.4542, and 2.5675 for substrates I, II, and III, respectively. The fundamental TE mode profile in each case is shown in Figure 1.

**Table 1.** Designed waveguide dimensions.

| Substrate | Si thickness (H) | Rib width (w) | Etch depth (h) |
|---|---|---|---|
| I | 5 μm | 5 μm | 2.25 μm |
| II | 3 μm | 2.25 μm | 1.75 μm |
| III | 220 nm | 500 nm | 130 nm |

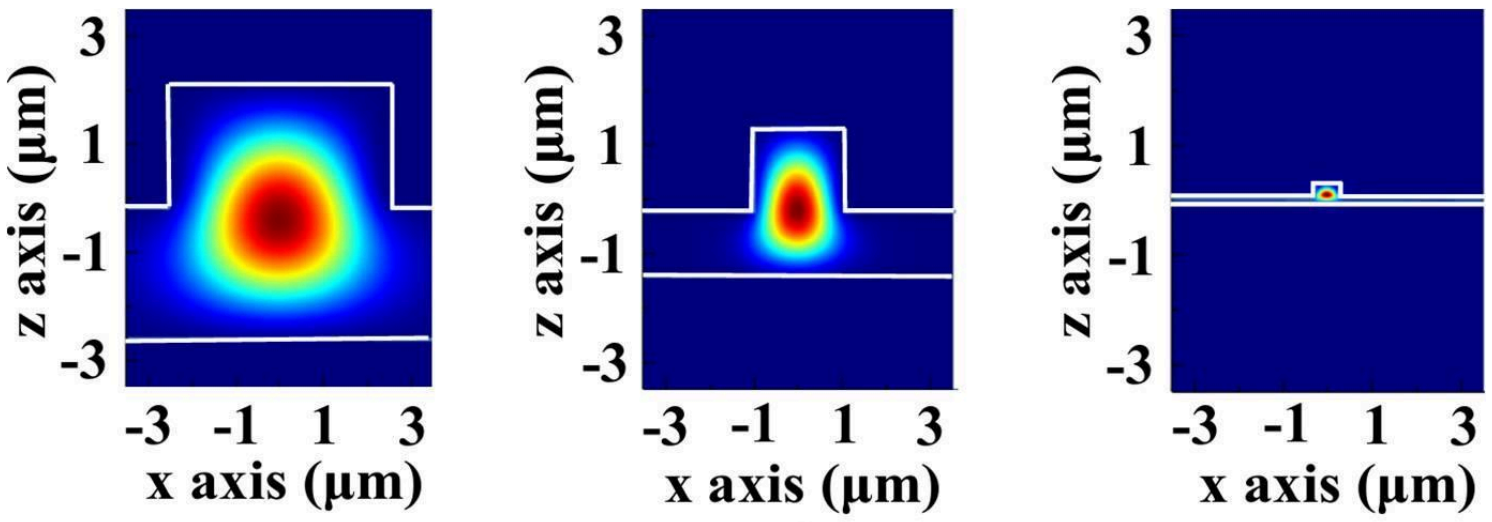

**Fig. 1.** Fundamental TE mode profiles for the single mode rib waveguides designed on (left) 5 μm, (middle) 3 μm, and (right) 0.22 μm SOI substrates.

## 2.2 Phase shifter optimization

After finding the single-mode waveguide design for TE polarization on different substrates, the next objective was to design phase modulators on these waveguides. The schematic of the phase shifter in a cross-sectional view, including the doped rib waveguides, is shown in Figure 2.

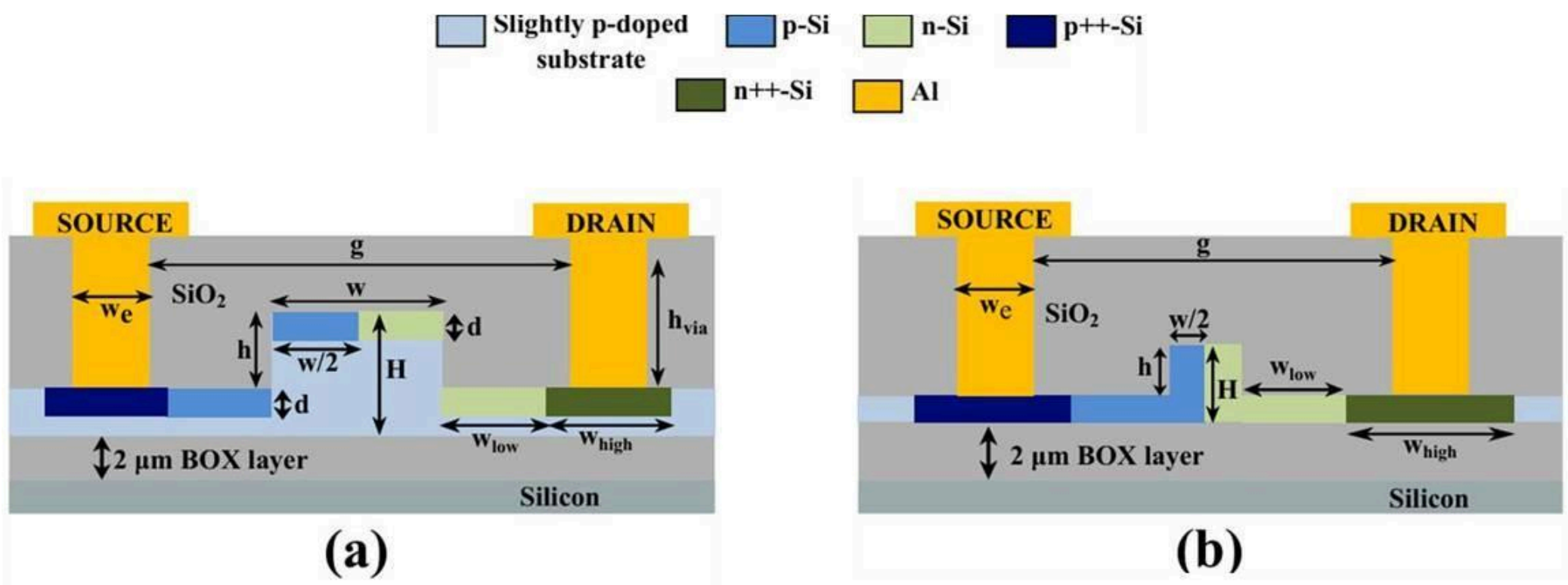

**Fig. 2.** Cross-sectional view of the phase shifter, that reflects source and drain electrodes, waveguide, and various doped regions using (a) thick film SOI substrates (5 μm and 3 μm SOI), and (b) thin film SOI (0.22 μm) substrates.

The thin-film SOI-based design (Figure 2(b)) uses a configuration selected from the established literature [11, 19, 21, 22], while the thick-film SOI-based design (Figure 2(a)) introduces a novel arrangement that incorporates a different doping profile from previous works [13, 23, 24]. Since intrinsic silicon does not have an electro-optic property inherently, to induce this property through the free carrier dispersion effect, we use trivalent/p-type (e.g., Boron) and pentavalent/n-type (e.g., Phosphorus) impurities [25]. As is the case in commercially available wafers, we start with a slightly p-doped device layer (light blue) with a doping concentration of $10^{15}/cm^3$, then moderate doping concentrations of p-type (sky blue) and n-type (light green) are considered for the formation of junctions. In particular, for thick films as shown in Figure 2(a), we choose doping depths, e.g., based on the feasibility through ion implantation, that cover only a part of the rib and slab regions, leading to the formation of p-i-n junctions. On the other hand, in the case of thin film substrates (Figure 2(b)), doping of the entire device layer is considered, creating p-n junctions. Highly doped $p$++ (deep blue) and $n$++ (deep green) regions in the slab with a doping concentration of $1 \times 10^{20}/cm^3$ act as ohmic contacts for placing the electrodes, which enhances the tunneling of carriers, enabling the formation of the p-i-n/p-n junctions[26]. When a reverse bias is applied to the junctions, it changes the carrier concentration of the device layer, which in turn changes the refractive index of the region. The carrier dispersion effect is defined by the relations given in Equations 1 and 2 [25],

$$\Delta n = -\frac{e^2\lambda^2}{8\pi^2 c^2 \varepsilon_0 n}\left(\frac{\Delta N_e}{m_e} + \frac{\Delta N_h}{m_h}\right) \qquad (1)$$

$$\Delta\alpha = \frac{e^3\lambda^2}{4\pi^2 c^3 \varepsilon_0 n}\left(\frac{\Delta N_e}{m_e^2 \mu_e} + \frac{\Delta N_h}{m_h^2 \mu_h}\right) \qquad (2)$$

where $\Delta n$ is the change in real part of refractive index, and $\Delta\alpha$ is the change in its imaginary part (i.e., extinction coefficient related to absorption). $e$ is the electronic charge; $\lambda$ is the operating wavelength; $\varepsilon_0$ is the permittivity of free space; n is the refractive index of the unperturbed material; $\Delta N_e$ and $\Delta N_h$ are the free carrier concentrations of electrons and holes, respectively; $m_e$ and $m_h$ are the relative masses of electrons and holes, respectively; $\mu_e$ and $\mu_h$ are the mobilities of electrons and holes, respectively; and $c$ is the speed of light. As can be seen from Equation (1), increasing the reverse bias reduces the free carrier concentrations in the junction, which in turn increases the refractive index of the material. Similarly, this effect causes the extinction coefficient in Equation (2) to reduce. To incorporate this carrier dispersion effect into our phase shifter design, we chose to optimize several key parameters, including doping concentration for moderately doped p and n regions, doping depth, electrode gap, and electrode width. It has been established that for maximizing the phase shifter performance, the p-doped regions should have a lower concentration and a larger area than the n-doped regions [27]. But since our goal is to compare the phase shifter performance based on SOI substrates of different thicknesses, we considered a symmetric doping profile (equal doping concentration and area) between p and n regions. The electrical simulations of the phase shifters were done using the CHARGE solver of Ansys Lumerical. The doping depths for moderate and highly doped regions along the rib and slab are considered to be the same, so all this doping can be done in a single fabrication step.

As mentioned previously, we vary the doping concentration, the doping depth ($d$), the electrode gap ($g$) and the electrode width ($w_e$) to obtain the maximum variation in the effective refractive index of the fundamental TE mode, $\Delta n_{eff}$, while maintaining a reasonable extinction coefficient. The doping concentrations of the moderately doped regions are varied from $1 \times 10^{17}/cm^3$ to $1 \times 10^{19}/cm^3$ in all three substrates. The resultant variations of $\Delta n_{eff}$ with respect to the applied reverse bias voltage are shown in Figure 3, where Figure 3(a,b,c) correspond to substrate I, II and III, respectively. These variations of $\Delta n_{eff}$ can be attributed to the change in charge carriers inside the guided mode volume. The nonlinear nature of the refractive index variations in all cases, although not clearly visible in Figure 3(a) because of scaling, is due to the carrier depletion effect producing a non-uniform distribution of carriers in the p-i-n /p-n junctions, leading to non-linear dependence of the depletion width on voltage [23]. The depletion width of the junction is given by,

$$w_D = \left( \frac{2\varepsilon_0 \varepsilon_r \left(V_0 + V\right)}{e} \left( \frac{1}{N_A} + \frac{1}{N_D} \right) \right)^{\frac{1}{2}} \tag{3}$$

where, $\varepsilon_0$ is the vacuum permittivity and $\varepsilon_r$ is the relative permittivity of silicon, $V_0$ is the built in potential, $V$ is the applied voltage, $N_A$ and $N_D$ are the concentrations of acceptors and donors, respectively; and $e$ is the electron charge.

For the thick-film substrates, in Figure 3(a,b), we focus on finding a doping concentration leading to the maximum $\Delta n_{eff}$, which occurs for a doping concentration $5 \times 10^{18}/cm^3$ for substrate I and $5 \times 10^{17}/cm^3$ for substrate II. In the case of substrate III, although higher doping concentrations give higher $\Delta n_{eff}$ values as shown in Figure 3(c), it also causes higher optical losses, as expected from Equation (4). In the thin film, since the entire device layer is doped as opposed to a partial covering in thick films, the overlap between the guided mode field and the free carriers is much higher, leading to larger refractive index variations as well as larger losses. Therefore, in the thin film, we have chosen a carrier concentration of $1 \times 10^{17}/cm^3$ as a trade-off for obtaining the required refractive index variation while maintaining reasonable carrier-induced optical losses (< 2 dB/cm) as shown in Figure 9(a)).

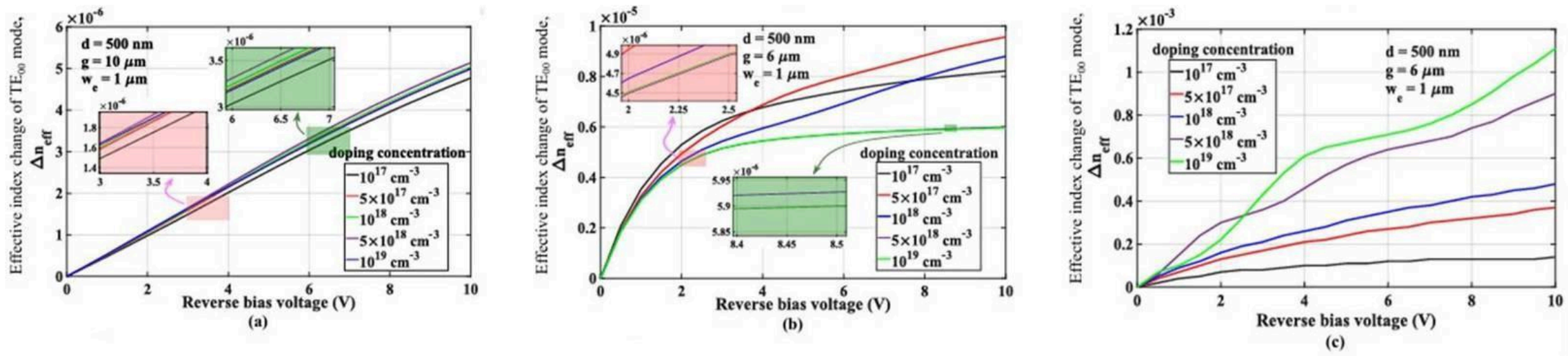

**Fig. 3.** $\Delta n$eff with reverse bias voltage for (a) 5 μm (substrate I), (b) 3 μm (substrate II), and (c) 0.22 μm (substrate III) substrates parametrized for various doping concentrations.

Similarly, the variation in the imaginary part ($\alpha$) of the effective index of the fundamental mode with the reverse bias voltage is shown for substrates I, II, and III in Figures 4(a,b,c), when subjected to different doping concentrations. The parameter $\alpha$ relates to the optical losses of the device as

$$loss = -\ 20log_{10}\left(e^{-\frac{2\pi\alpha z}{\lambda_0}}\right) \tag{4}$$

where, z is the propagation length. It can be seen that the values of $\alpha$ show a decreasing trend with increasing reverse bias. The reason is that, while subjected to depletion, free carrier absorption decreases due to the fewer carriers left, which causes the $\alpha$ value to reduce. The same effect is also observed with respect to the variation in doping concentration as a result of the availability of lesser free carriers at lower doping concentrations.

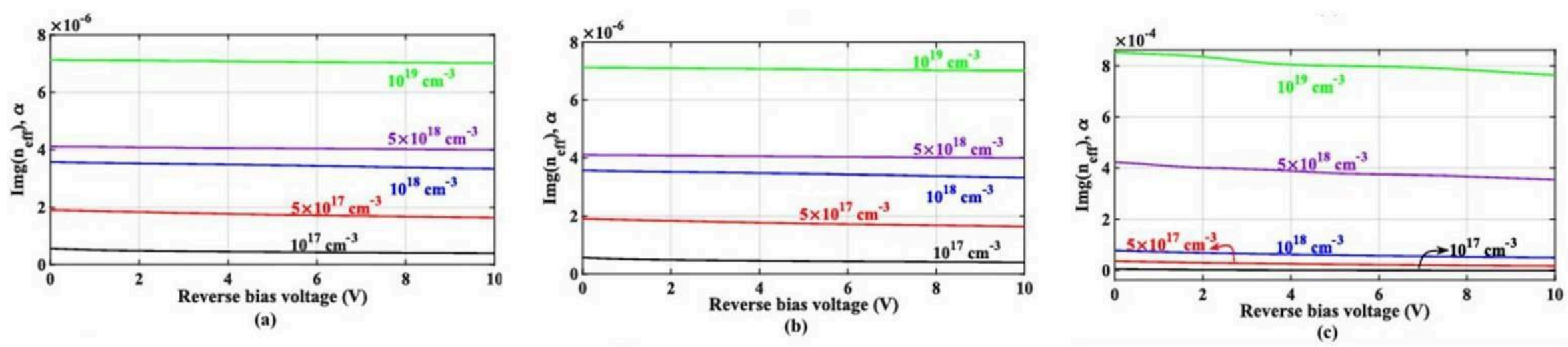

**Fig. 4.** Variation of the imaginary part of the effective index, $\alpha$, with respect to reverse bias voltage for (a) 5 µm (substrate I), (b) 3 µm (substrate II), and (c) 0.22 µm (substrate III) substrates, parametrized for various doping concentrations.

Regarding doping depth optimization, the entire device layer of the thin film substrate (III) is doped, as is common practice. In contrast, for thick film substrates, the doping depth is varied from 100 nm to 500 nm. A larger doping depth gives better interaction between the optical mode field and the charge carriers, leading to a higher $\Delta n_{\text{eff}}$ as can be seen in Figure 5(a) and Figure 5(b) for substrates I and II, respectively. Subsequently, the variation of $\alpha$ is presented in Figure 6(a) and Figure 6(b) for different doping depths in substrate I and II, respectively, under reverse bias conditions. As expected, a larger doping depth causes $\alpha$ to have a larger value, thus enhancing optical absorption.

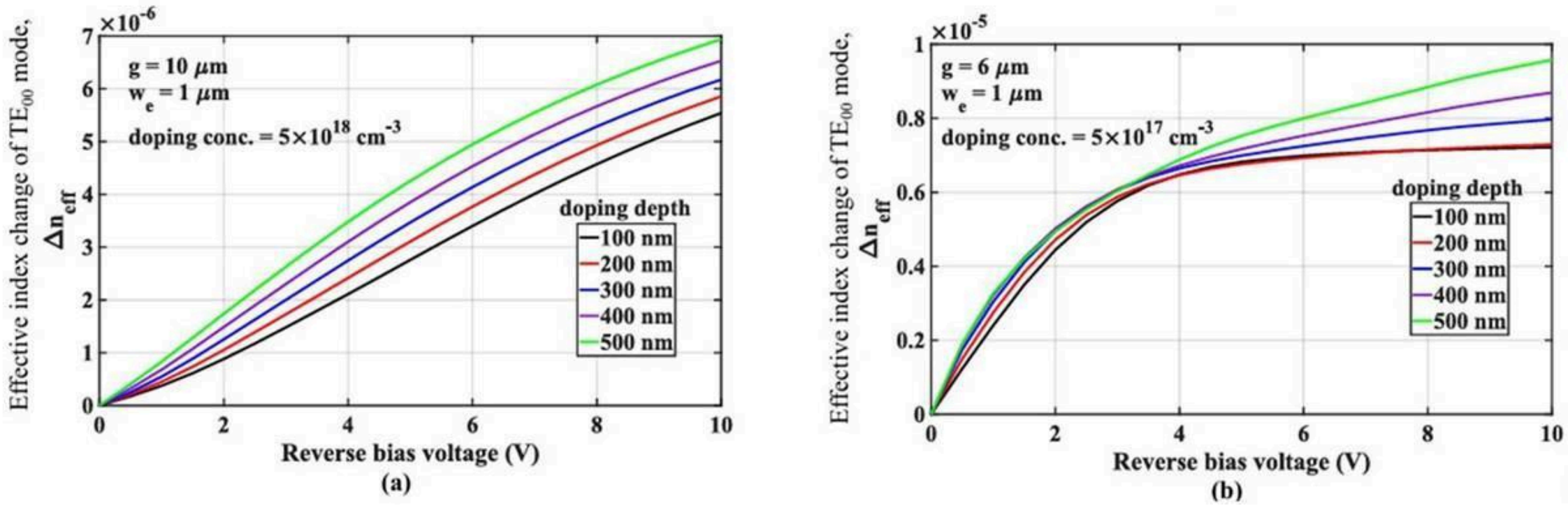

**Fig. 5.** $\Delta n$eff with reverse bias for (a) 5 µm (substrate I), and (b) 3 µm (substrate II) substrates parametrized for varying doping depths.

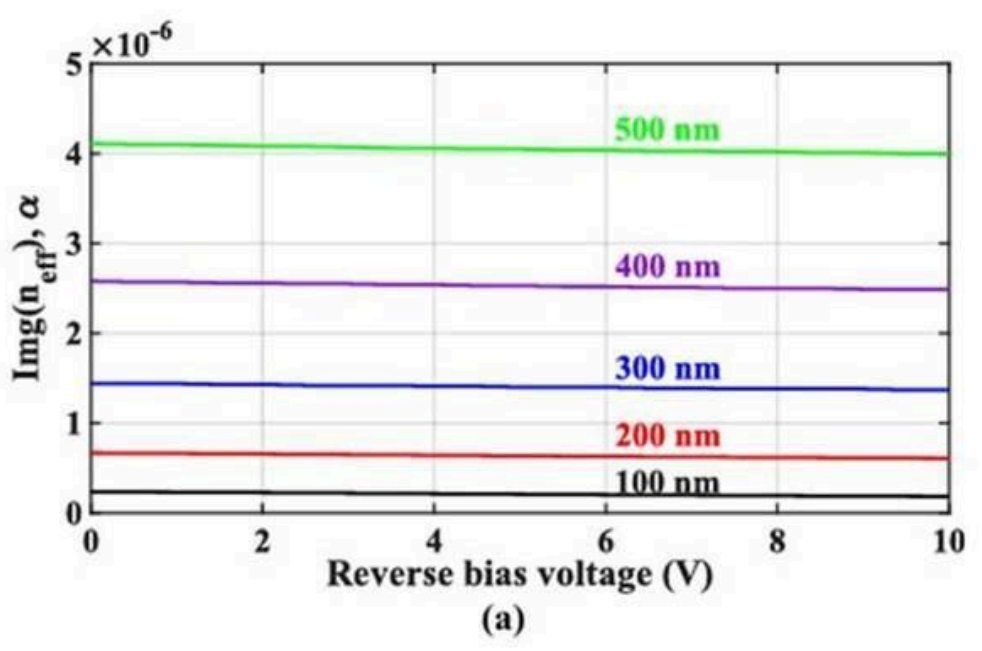

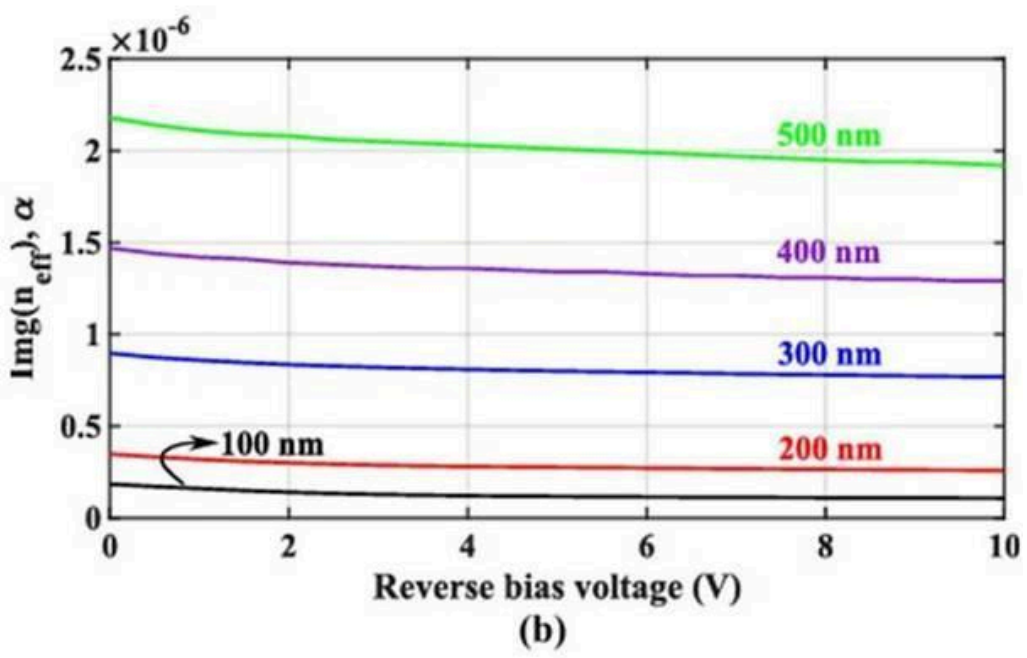

**Fig. 6.** Imaginary part of the effective index, $\alpha$ with respect to reverse bias voltage for (a) 5 µm (substrate I), and (b) 3 µm (substrate II) substrates parametrized for various doping depths.

The electrode gaps are varied from 6 µm to 10 µm for substrate I, 4 µm to 8 µm in case of substrate II, and 4 µm to 6 µm in case of substrate III. We have chosen these different ranges for different substrates keeping in mind the difference in waveguide widths in each substrate as required for their single-mode behavior (Figure 1). Additionally, the minimum gap is set by taking into account the possibility of non-vertical sidewalls of the waveguides. Figure 7(a) and Figure 7(b) show the $\Delta n_{\text{eff}}$ for different electrode gaps in case of substrates I and II, respectively. It can be seen that as the electrode gap reduces, $\Delta n_{\text{eff}}$ increases. Henceforth, the electrodes should be kept closer to the waveguide edges, but not too close to touch the evanescent field to avoid further optical losses. Consequently, the electrode gaps have been chosen as 8 µm for substrate I and 5 µm for substrate II, considering the corresponding extinction coefficients ($\alpha$) from Figure 8(a) and Figure 8(b). As observed, a smaller electrode gap leads to free carriers being confined closer to the optical mode field, which increases the free carrier-optical mode overlap, thereby increasing absorption. In addition, closer electrodes intensify local electric field, which enhances the separation and distribution of charge carriers by exerting stronger electrostatic forces. Substrate III experiences negligible change in $\Delta n_{\text{eff}}$ with variation of electrode gap, and remains same as that in Figure 3(c). Similarly, electrode width is varied from 1 µm to 2 µm. Within this range, there was only a slight change in $\Delta n_{\text{eff}}$ for substrates I and II, and almost no change for substrate III. Hence, the electrode gap and the electrode width for substrate III are chosen such that the modal overlap with the charge carriers were minimized [12]. Based on this analysis, suitable electrode widths were chosen for all the three substrates, along with an electrode gap of 6 µm for substrate III. On the basis of the analysis of various parameters mentioned above, we compare the performance of phase-shifter designs in the three substrates. In particular, we get $\Delta n_{\text{eff}}$ of the fundamental TE modes as $4.3 \times 10^{-6}$, $7.4 \times 10^{-6}$, and $1.1 \times 10^{-4}$ on substrates I, II, and III, respectively, through the optimization of various parameters as detailed in the previous section, while the reverse bias voltage varies from 0 V to 5 V. The corresponding $V_{\pi}L$ for the three devices are given in Table 2. It can be seen from the values that thick film-based designs require either a larger phase shifter length or higher reverse bias voltages to achieve the $\pi$ phase shift for the full modulation depth, whereas for the thin film, both the required voltage and the length are within the desired range. Therefore, for thick films, there must be a trade-off between a CMOS-compatible voltage, $V_{\pi}$, and a feasible length for the standard fabrication process. For example, for the thin film, if we consider a $V_{\pi}$ of 5 V, then the phase shifter length would be 6.45 mm. The optical losses for the three phase shifter designs, as well as the relative phase shift for a phase shifter length of 6.45 mm, are shown in Figure 9. In Figure 9(c), the phase changes are not linearly dependent on the voltage due to the non-linear dependence of the depletion width on the voltage [23], as given in Equation (3).

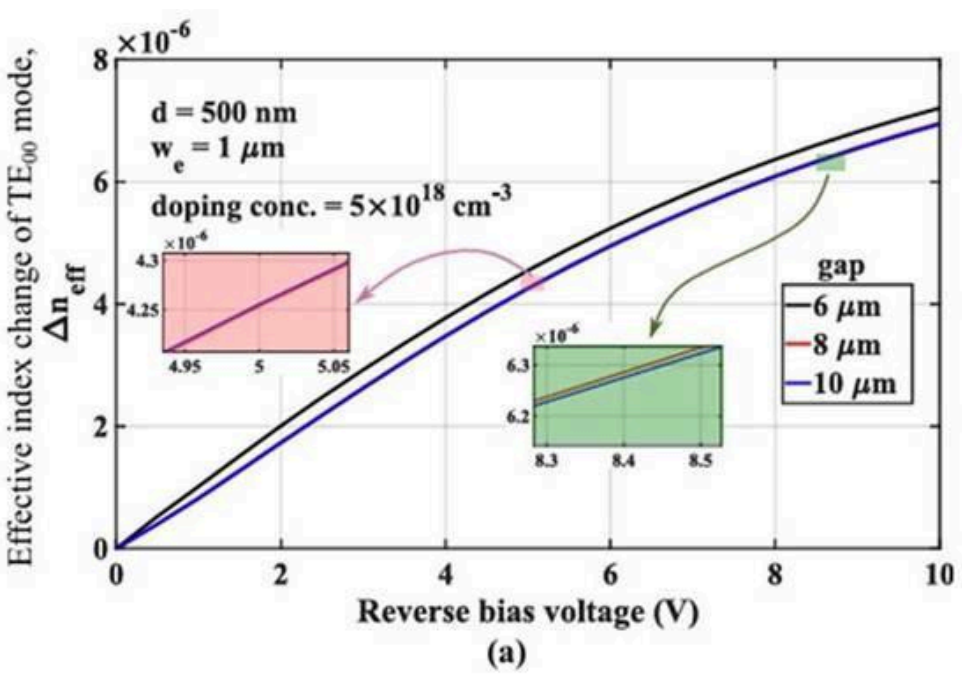

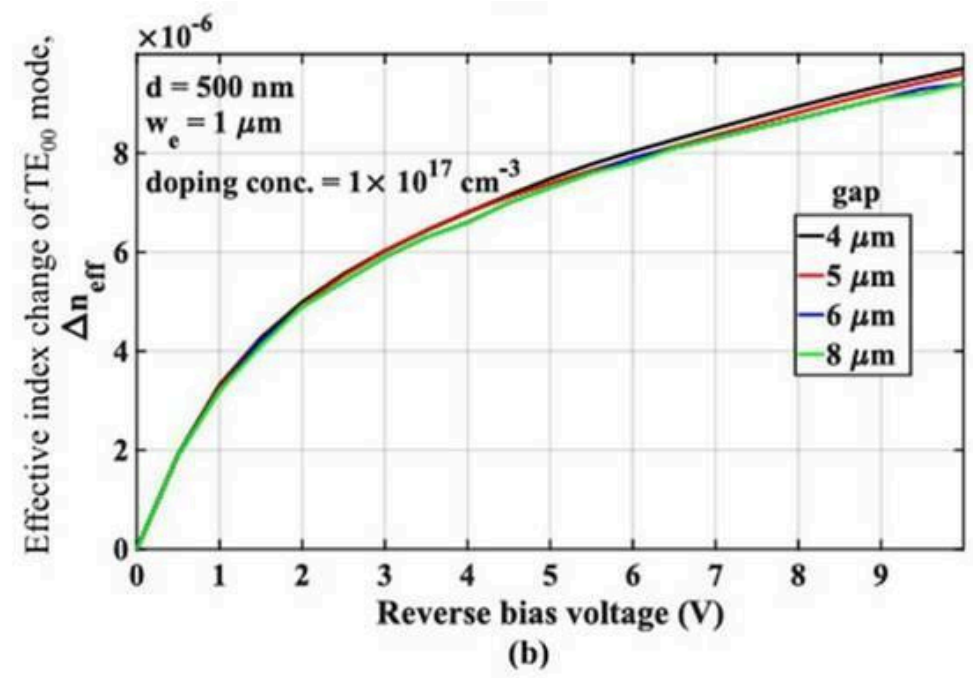

**Fig. 7.** Δ*n*eff with reverse bias (a) 5 μm (substrate I), and (b) 3 μm (substrate II) substrates parametrized for varying electrode gaps.

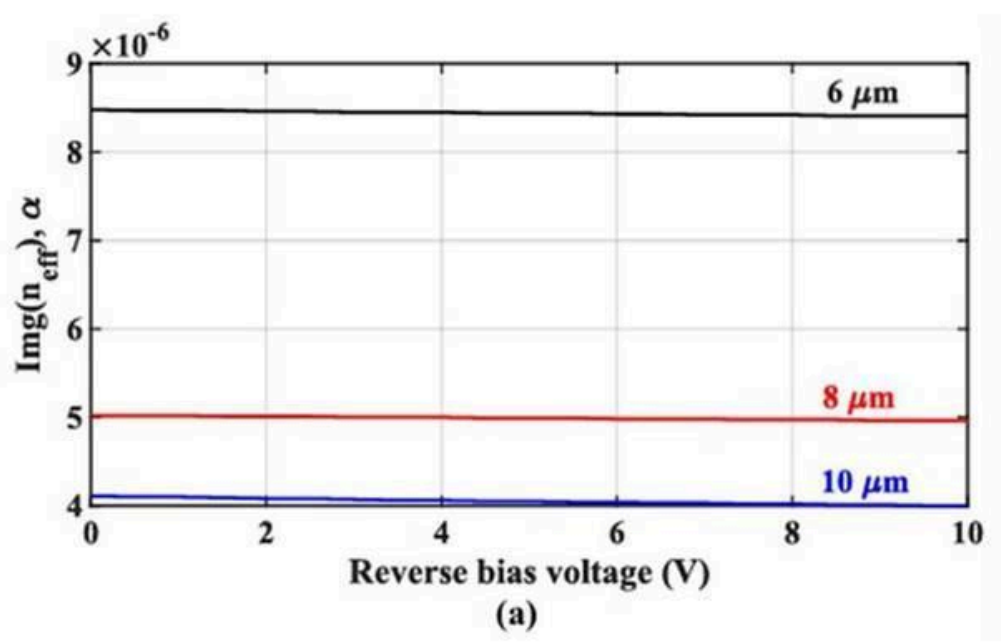

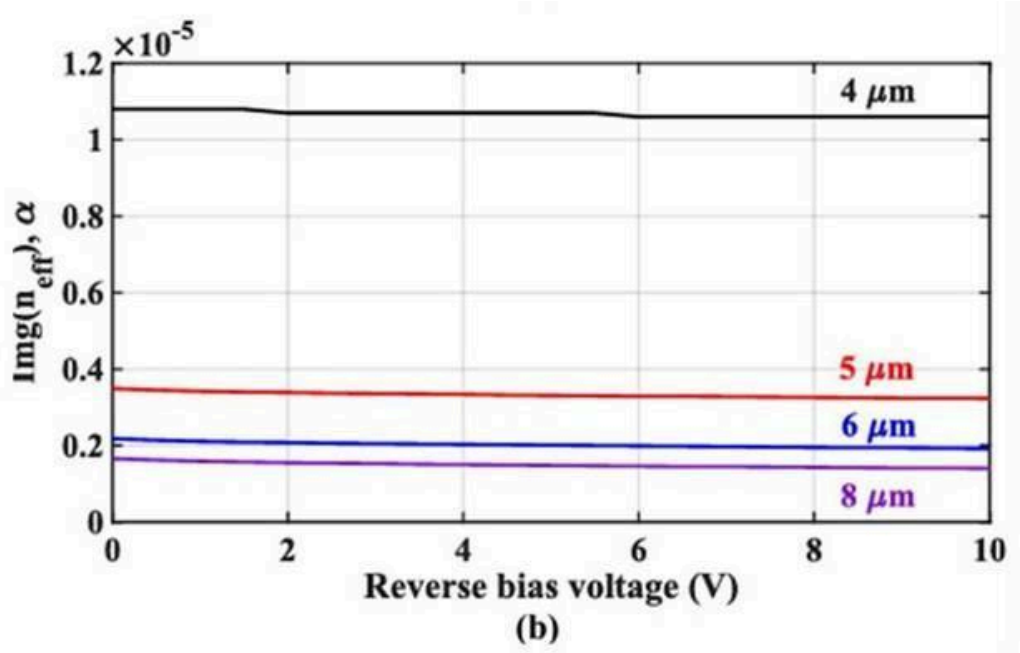

**Fig. 8.** Imaginary part of the effective index, $\alpha$ with respect to reverse bias voltage for (a) 5 μm (substrate I), and (b) 3 μm (substrate II) substrates parametrized for varying electrode gaps.

**Table 2.** $V_\pi L$ values for the phase shifters.

| Substrate | $V_\pi L$ |
|---|---|
| I | 88 V·cm |
| II | 4.5 V·cm |
| III | 3.17 V·cm |

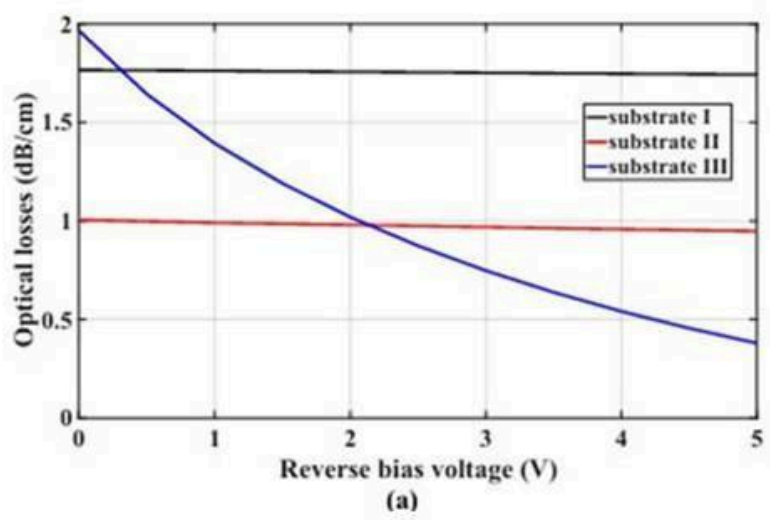

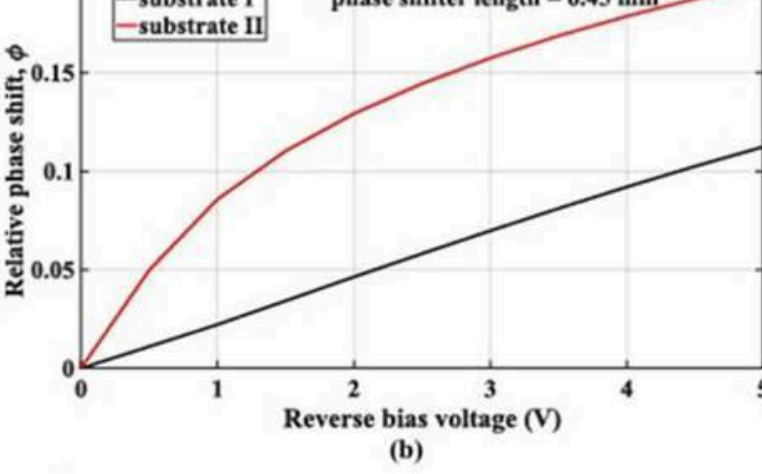

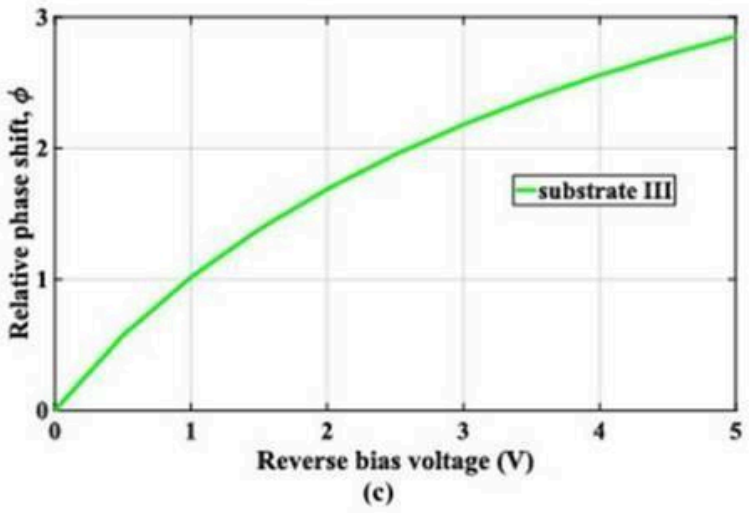

**Fig. 9.** (a) Optical losses with respect to reverse bias voltage; (b) relative phase shift for phase shifter designs on substrates I, II, and (c) III.

### 2.3 Mach-Zehnder modulator (MZM) operation and performances

To validate the above phase-shifter designs, we configured intensity modulators on all three SOI substrates using the phase shifter in one of the arms of a Mach-Zehnder Interferometer (MZI). Since we are conducting a comparative study, we chose a phase-shifter length of around 6.45 mm in all three substrates, although it provides a $V_\pi$ of 5 V only for the thin-film substrate as mentioned in the previous section. The MZM schematic is shown in Figure 10.

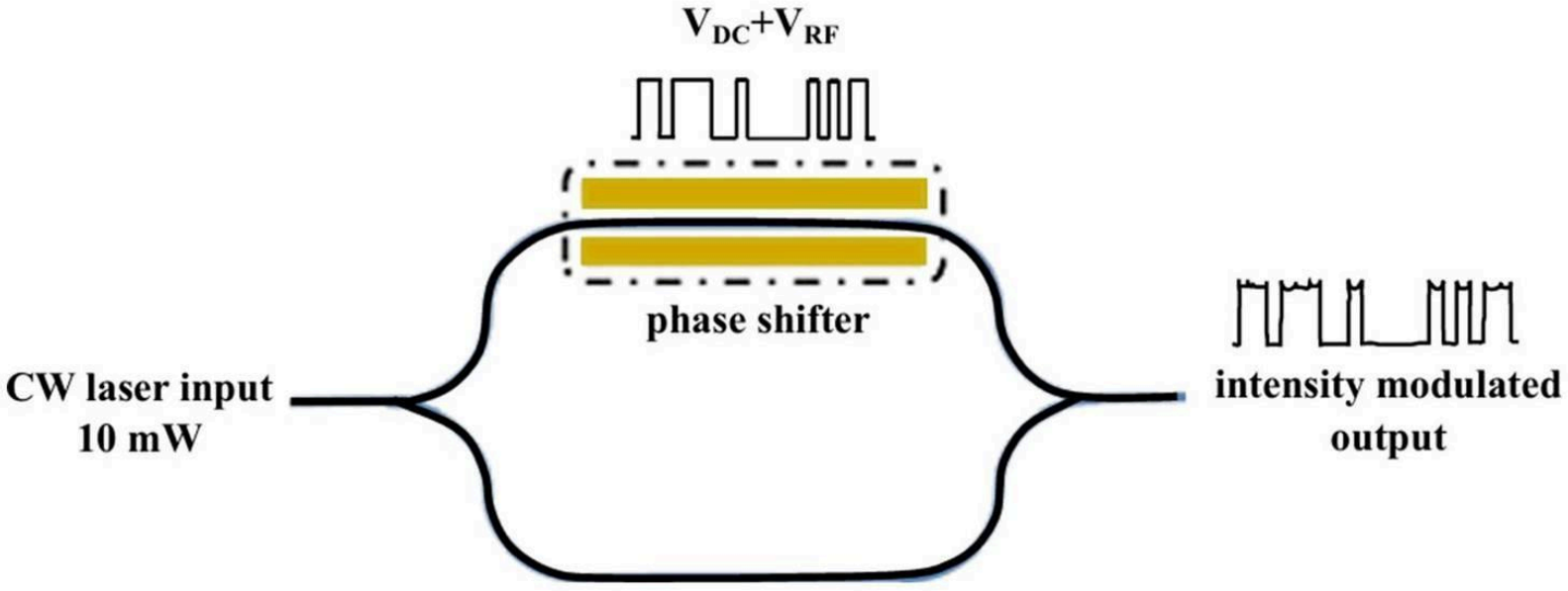

**Fig. 10.** Schematic of the MZM having a phase shifter in one arm.

The electrodes are placed on the slab region, spanning the entire length of the phase-shifter. Depending on the radio-frequency (RF) signal applied, the electrodes can be lumped or traveling wave type. Suppose the electrode length is sufficiently smaller (one-tenth) compared to the RF wavelength (i.e, $\lambda_{RF} < \lambda_0/n_R$ ). In that case, it is a lumped electrode configuration, and greater in the case of a traveling-wave configuration [28]. Here, $\lambda_0$ is the RF wavelength in vacuum, and $n_{\mathrm{RF}}$ is the RF index. Two Y-splitters/combiners are used to split/combine the input continuous wave (CW) signals into the two arms, which then interfere to produce the modulated output. We consider an input signal power of 10 mW to diagnose the signal modulation with less jitter and amplitude distortion. The output intensity $I$ of an MZM is given as

$$I = \frac{I_0}{2}[1 + cos\Delta\phi] \tag{5}$$

where $I_0$ is the input intensity and $\Delta\phi$ is the phase change. The Y-splitter/combiner element is simulated using the Ansys Lumerical INTERCONNECT module, providing an insertion loss of 2.6 dB [29]. Aluminum is used for the vias as well as contact pads to which both direct current (DC) bias and radio frequency (RF) signals can be applied. An upper cladding of silicon dioxide ($SiO_2$) was considered, which encapsulates the waveguide layer just enough to expose the contact pads. For simulation, we have chosen a via depth ($h_{\mathrm{via}}$) equal to the rib height, and contact pad thickness of 2 µm with a width ($w_{\mathrm{e}}$) of 50 µm. The corresponding RF losses and the RF group indices are shown in Figure 11 for the three substrates. It can be seen from Figure 11(a) that RF losses start increasing rapidly above 1 GHz for all the substrates, which can be further improved by careful design of the RF transmission lines. It can also be observed from Figure 11(b) that the mismatch between the optical and RF indices for the modulators is reduced close to their electro-optic cut-off frequency. At frequencies higher than 1 GHz, we get a close match between the RF and guided mode indices, which is critical for high-frequency modulator design [30], in particular, because in this frequency range, the RF losses typically start increasing. At low frequencies, though, the index mismatch between the RF and the guided mode is higher, its impact is less critical since, in addition to the very low RF losses, the modulation mechanism differs in this range.

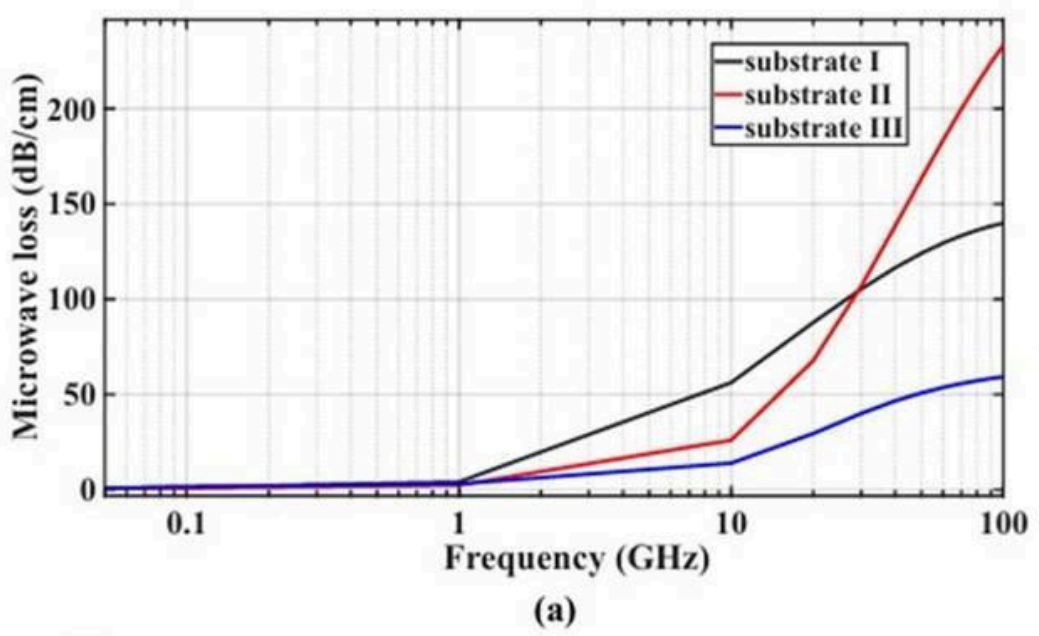

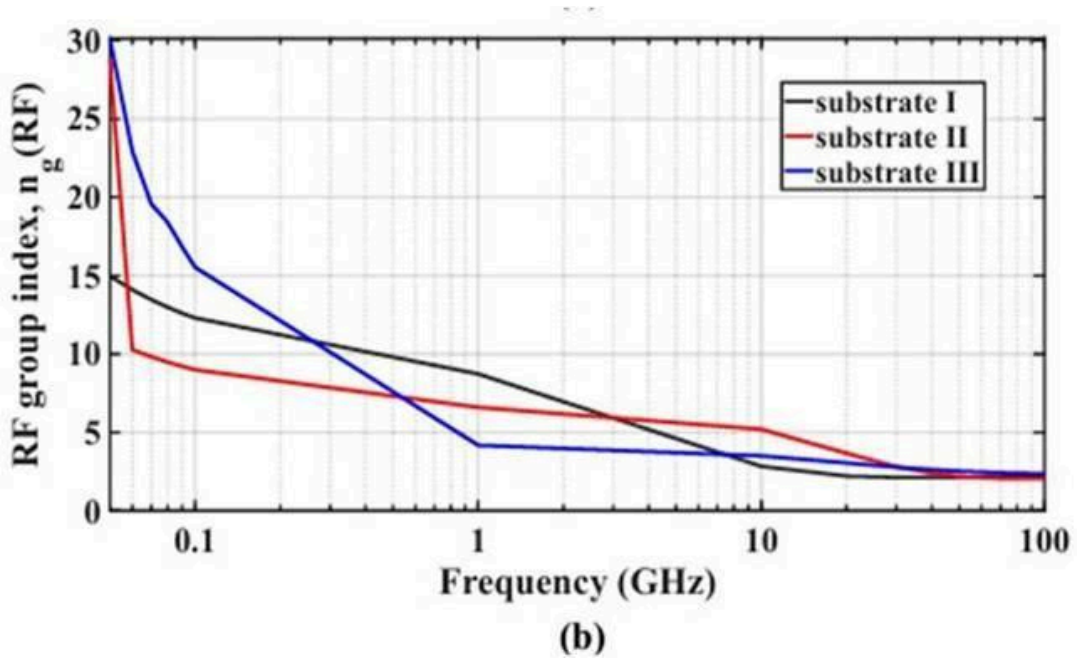

**Fig. 11.** (a) RF losses with respect to frequency, and (b) group index for intensity modulator designs on substrates I, II, and III.

### 2.3.1 Static characteristics

To validate the system-level performance of the designed phase shifters in the MZM, we simulated static characteristics using Lumerical INTERCONNECT. This step is essential because it provides insight into the phase shifter's behavior under various bias conditions, helping to verify its phase tuning range, losses, and wavelength shift. Figure 12 shows the response spectra of the MZMs with different driving voltages. For all the substrates, we choose a length of ~ 6.45 mm, which corresponds to a full modulation only for substrate III for a bias of 5 V (Table 2). An extra length of 100 μm is introduced into the lower arm of the MZM (Figure 10), to simulate the transmission spectra. However, it is worth noting that this value of length mismatch is used only to determine optical transmission characteristics with respect to different biases and is not applied in the actual MZM. At 0 V ($\Delta\phi = 0$), the transfer function in Equation (5) indicates the maximum intensity, which gives the static extinction ratio of the devices. It can be seen from the figures that an increase in reverse bias causes a left shift in the wavelength, as well as changes the intensity level of the transmitted signal according to the transfer function of the respective modulators. The wavelength shift in response to the bias voltage is lowest in the case of substrate I (Figure 12(a)) and highest for the substrate III-based modulator (Figure 12(c)), which is a direct indication of their modulation depth. As can be seen in the figures, the achieved static extinction ratio values are less than 20 dB, which is typical for silicon modulators [31]. In a practical and unbalanced MZI, losses in the arms (due to absorption and imperfect splitting) will broaden the spectral features, leading to lower fringe contrast and wider resonance-like notches. The resonance notches for thick film substrates are narrower, indicating a higher Q-factor, which is attributed to their lower losses as discussed in section 2. In contrast, the thin-film substrate-based device exhibits a broader tuning range but with wider resonance peaks, which corresponds to a lower Q-factor due to higher losses. The static characteristics show the output intensities of the non-ideal modulator, taking absorption coefficients into consideration. The ideal transfer function given in Equation 5 no longer satisfies the behavior of the output intensity.

$$I = \frac{I_0}{4}\left[\tau_1^{\ 2} + \tau_2^{\ 2} + 2\tau_1\tau_2 cos\Delta\phi\right] \qquad (6)$$

where, $\tau_i^{\ 2} = exp(-\alpha_i L_i)$. $\alpha_i$ is the extinction/absorption coefficient as shown in Section 2.2, in each arm, $L_i$ is the length of each arm, and $\Delta\phi$ is the phase difference between the two arms.

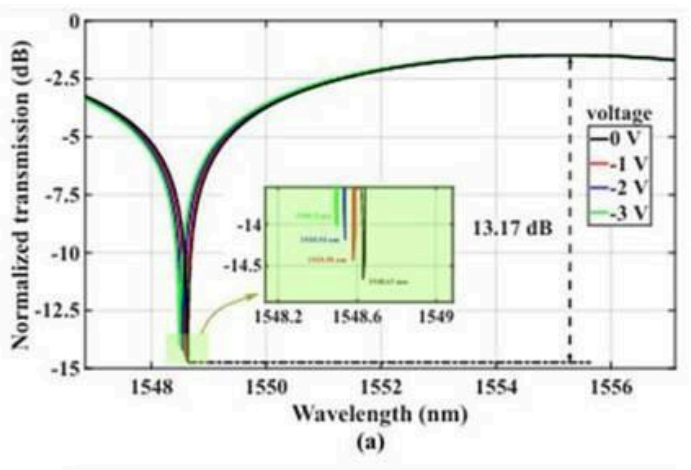

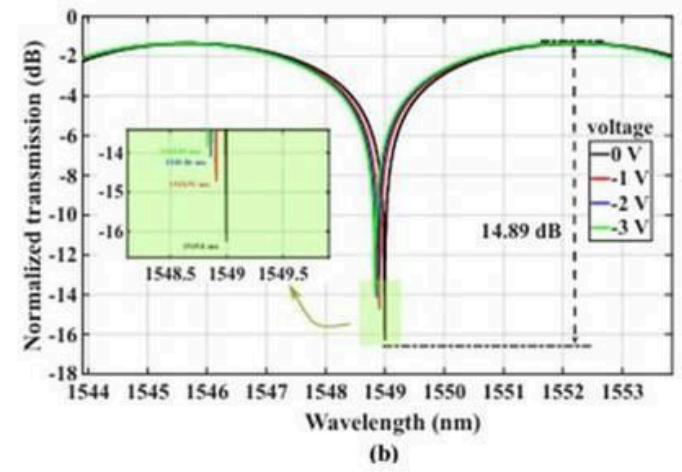

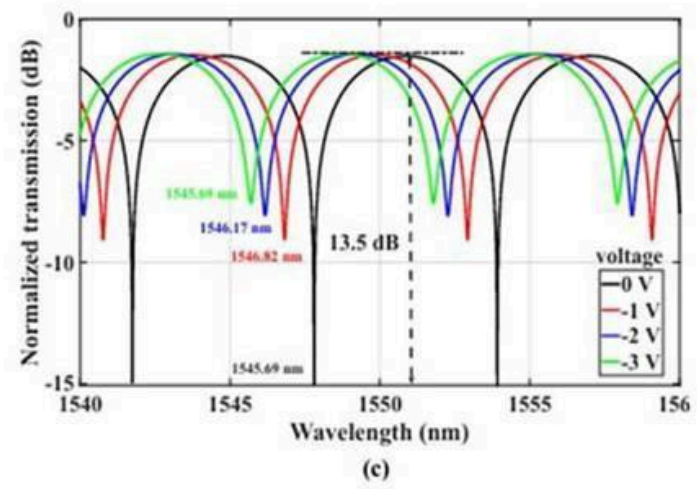

**Fig. 12.** The normalized transmission spectra of the modulators based on (a) substrate I, (b) substrate II, and (c) substrate III with different reverse bias voltages.

### 2.3.2 Bandwidth estimation

Considering a load resistance of 50 Ω, the RC limited bandwidths of the devices are found to be 3.6 GHz, 8.9 GHz, and 12.9 GHz for substrates I, II and III, respectively. These estimates are calculated using the slab resistance and junction capacitance obtained for the three devices. The slab resistance values for substrates I, II, and III are 22731.3 Ω·μm, 17154.1 Ω·μm, and 56495.7 Ω·μm, respectively. The corresponding junction capacitance values are 1.93086 fF/μm, 1.03638 fF/μm, and 0.219631 fF/μm. The 3 dB electro-optic bandwidth of the designed MZMs obtained from the $S_{21}$ parameters in the INTERCONNECT simulations, as shown in Figure 13, are 3.5 GHz, 6.5 GHz, and 9.5 GHz respectively for substrates I, II, and III. The deviation of the bandwidth values achieved from system-level simulations in the INTERCONNECT module from the theoretically estimated values, particularly for the substrates II and III, stems from the underlying electrode configuration. Specifically, the INTERCONNECT simulation uses lumped electrode configurations for substrates I within their bandwidth limit, whereas for substrates II and III, a traveling-wave electrode configuration is considered near the cut-off frequency. The traveling wave electrode configuration is more energy intensive, leading to a lower bandwidth in practice [32].

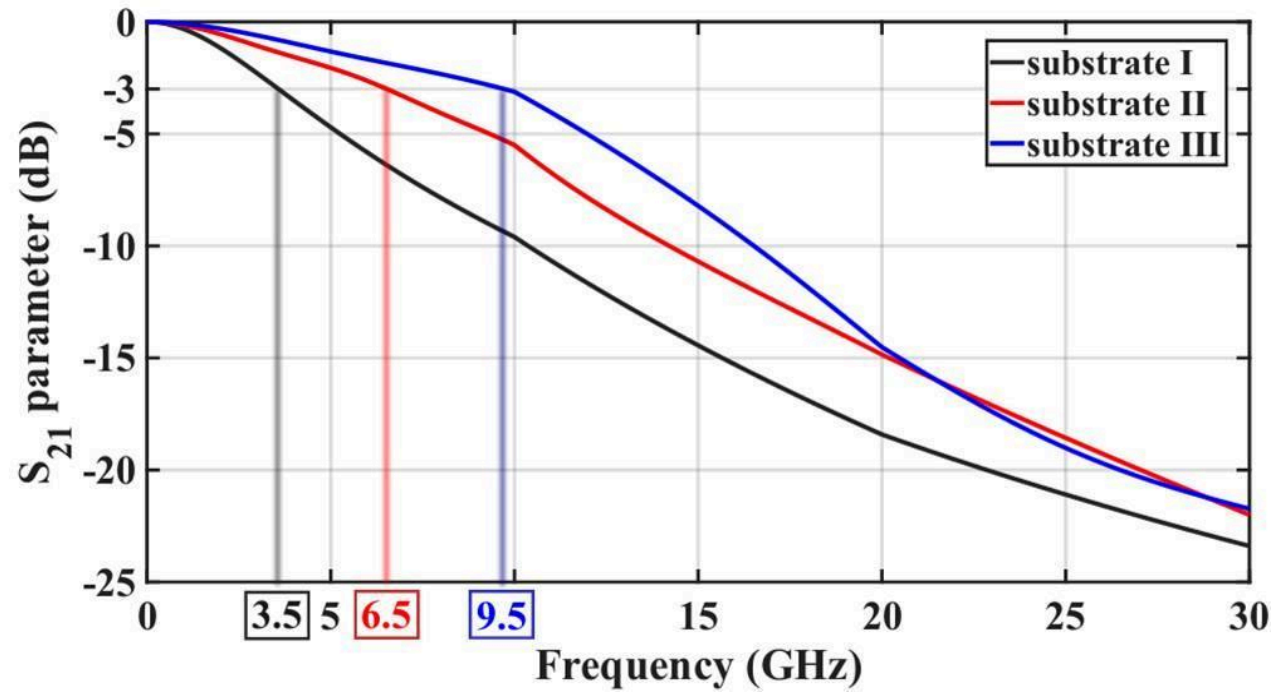

**Fig. 13.** Gain parameter of the modulators indicating the 3 dB bandwidths.

### 2.3.3 RF transmission characteristics

Based on the bandwidth estimation, we characterize the RF transmission of the MZMs for on-off-keying (OOK) modulation through the eye diagram analysis for non-return-to-zero (NRZ) pseudo-random binary sequences (PRBSs). The eye diagrams are shown in Figures 14, 15, and 16 for substrates I, II, and III, respectively, at frequencies above and below the cut-off frequencies in all three substrates. Note that we have carefully balanced the length difference between the two arms of the MZM to ensure that the modulator setpoint is close to the quadrature point. One can also use the DC bias voltage to fine-tune the quadrature setpoint. However, for the purpose of comparison, we set a fixed DC bias at 0 V for all three cases, and chose a suitable $V_{RF}$ below 5 V for near-perfect eye crossing.

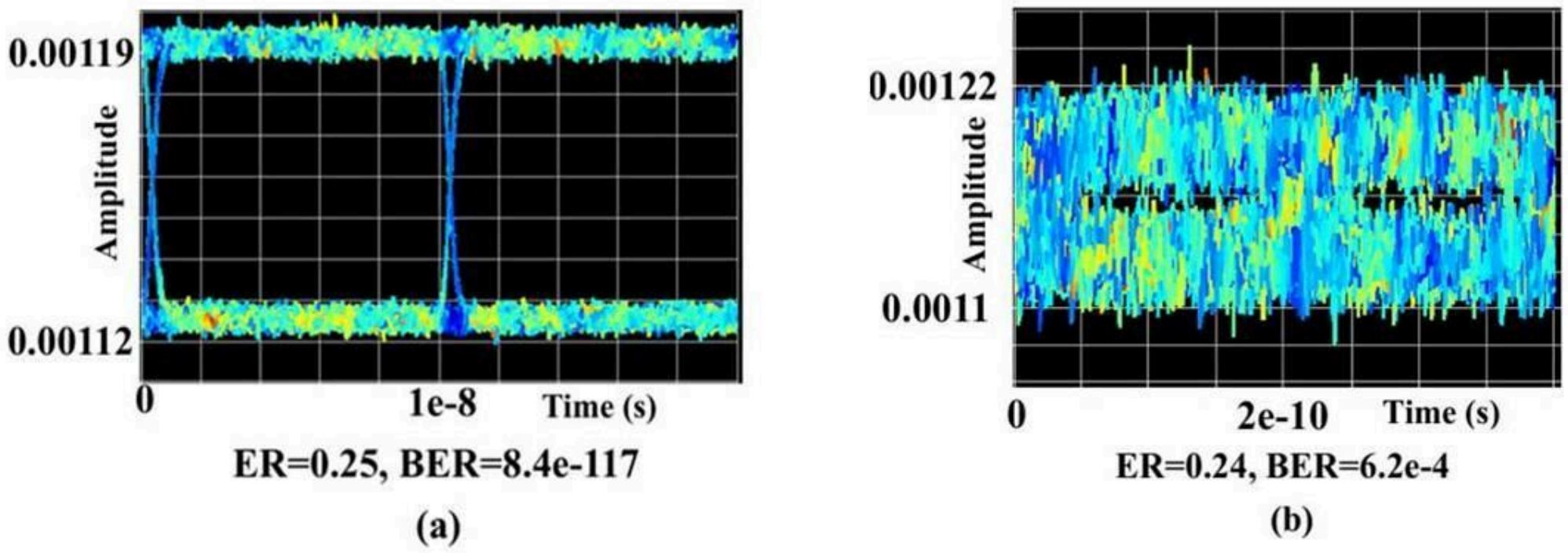

**Fig. 14.** Intensity modulation performance for 5 μm SOI based MZM, at (a) 100 Mbps, and (b) 5 Gbps, for RF signal of 2.5 V.

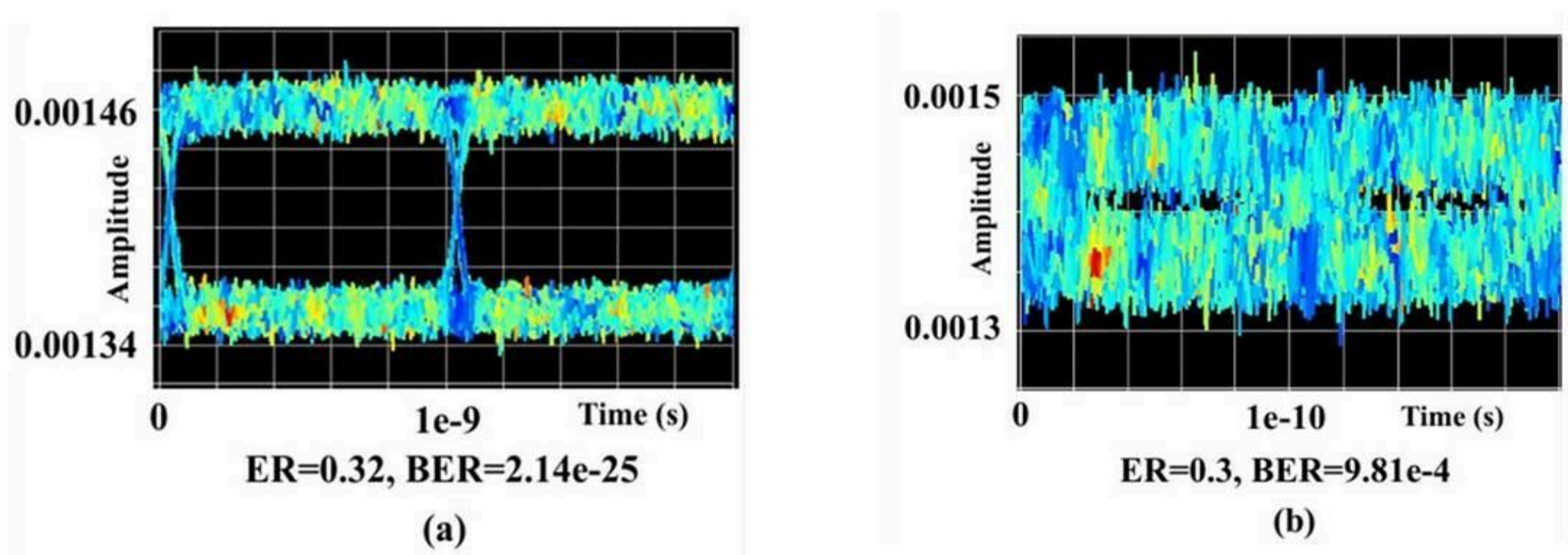

**Fig. 15.** Intensity modulation performance for 3 μm SOI based MZM, at (a) 1 Gbps, and (b) 10 Gbps, for RF signal of 0.8 V.

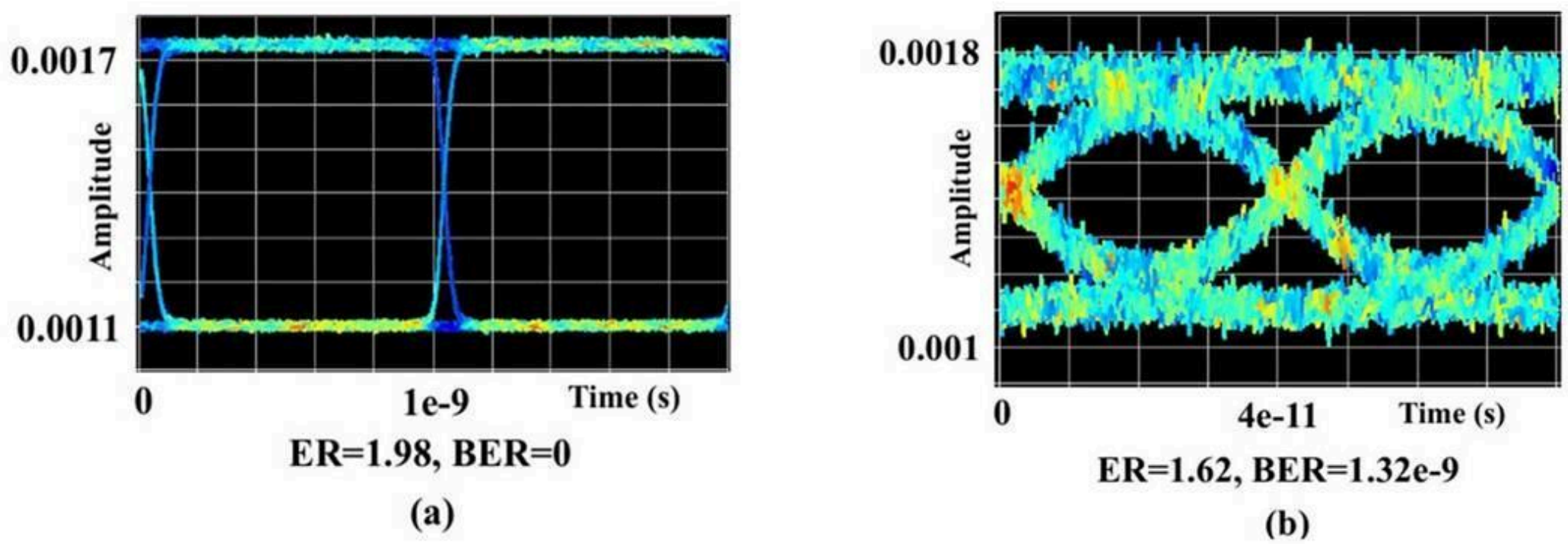

**Fig. 16.** Intensity modulation performance for 220 nm SOI based MZM, at (a) 1 Gbps, and (b) 20 Gbps, for RF signal of 0.5 V.

For all three substrates, we observe clear eye-openings at signal frequencies well below their cut-off values, leading to extremely low bit error rates (BER). On the contrary, for signal frequencies higher than the corresponding cut-off frequencies, the eye-opening significantly degrades, indicating higher time jitter, intersymbol interference, and BER. Moreover, the extinction ratio increases for the thinner substrates due to their enhanced modulation efficiency.

### 3. Conclusion

This work provides a comprehensive overview of carrier-depletion-based MZMs designed on commercially available SOI substrates with different device-layer thicknesses. In summary, the presented design approach reveals that 5 μm SOI substrates are optimal for applications prioritizing low propagation losses (< 1.8 dB/cm) and moderate operating speeds on the order of several hundred megahertz. 3 μm SOI substrates in the chosen design can accommodate even lower losses (< 1.5 dB/cm) while enabling operation into the multi-gigahertz range. For scenarios where higher speeds are essential, 220 nm SOI substrates facilitate operation at several tens of gigahertz, with associated losses maintained < 2 dB/cm. However, further optimization of thin-film-based designs could enable significantly greater speed enhancements, though this is generally accompanied by increased optical losses. To conclude, thin film substrates are better for applications involving high bandwidth requirements. However, thick film SOI-based MZMs may still find use in low-speed, low-integration density applications, especially where low losses and high power handling are required. They are also economical, in addition to providing higher fabrication tolerance, and ease of batch processing by means of optical lithography, whereas the thin-film-based devices require higher precision technology like electron beam lithography (EBL) or deep ultra-violet (DUV) lithography. Therefore, this study may provide users with flexibility on substrate choice, allowing the device structure to be optimized for specific system needs and targeted performance outcomes.

## Declarations

**Acknowledgements:** The authors thank Mr. Mijanur Rahaman for helping in resolving simulation challenges during the research.

**Research ethics:** Not applicable.

**Author contributions:** All the authors have accepted responsibility for the entire content of this submitted manuscript and approved submission.

**Competing interests:** The authors declare no conflicts of interest regarding this article.

**Research funding:** None declared.

**Data availability:** The raw data can be obtained on request from the corresponding author.

## References

[1] Siew SY, Li B, Gao F, Zheng HY, Zhang W, Guo P, et al. Review of silicon photonics technology and platform development. Journal of Lightwave Technology. 2021;39(13):4374-89.

[2] Preston K, Chen L, Manipatruni S, Lipson M. Silicon photonic interconnect with micrometer-scale devices. In: 2009 6th IEEE International Conference on Group IV Photonics. IEEE; 2009. p. 1-3.

[3] Aalto T, Cherchi M, Harjanne M, Bhat S, Heimala P, Sun F, et al. Open-access 3-$\mu$m SOI waveguide platform for dense photonic integrated circuits. IEEE Journal of selected topics in quantum electronics. 2019;25(5):1-9.

[4] Georgiev GV, Cao W, Zhang W, Ke L, Thomson DJ, Reed GT, et al. Near-IR & mid-IR silicon photonics modulators. Sensors. 2022;22(24):9620.

[5] Deng H, Zhang J, Soltanian E, Chen X, Pang C, Vaissiere N, et al. Single-chip silicon photonic engine for analog optical and microwave signals processing. Nature Communications. 2025;16(1):5087.

[6] Shu X, Cheng Z, Ma L, Chen B, Li C, Sun C, et al. On-chip silicon electro-optical modulator with ultra-high extinction ratio for distributed optical fiber sensing. In: Optical Fiber Sensors. Optica Publishing Group; 2023. p. Th5-4.

[7] Wu Y, Chu T. Ultralow-Crosstalk Silicon Electro-Optic Switch with Cascaded Phase Shifters for Loss Equivalence. arXiv preprint arXiv:241118139. 2024.

[8] Bao P, Yao C, Penty RV, Cheng Q. Silicon Electro-optic Mach-Zehnder Switch Fabric with Ultralow-crosstalk. Journal of Lightwave Technology. 2025.

[9] Ding Y, Zhang D, Zhang P, Tang B, Wang F, Wang X, et al. Dual-mode 2× 2 electro-optic switch on a SOI platform. Optics Letters. 2024;49(21):6125-8.

[10] Montesinos-Ballester M, Deniel L, Koompai N, Nguyen THN, Frigerio J, Ballabio A, et al. Mid-infrared integrated electro-optic modulator operating up to 225 MHz between 6.4 and 10.7 $\mu$m wavelength. ACS photonics. 2022;9(1):249-55.

[11] Mohammadi A, Safarnejadian A, Zheng Z, Zeng M, Rusch LA, Shi W. High-bandwidth silicon traveling-wave modulator using distributed micro-capacitors. Journal of Lightwave Technology. 2024;42(17):5942-8.

[12] Yan Y, Zhang H, Liu X, Peng L, Zhang Q, Yu G, et al. Recent Progress in Electro-Optic Modulators: Physical Phenomenon, Structures Properties, and Integration Strategy. Laser & Photonics Reviews. 2025;19(3):2400624.

[13] Hewitt P, Reed G. Improving the response of optical phase modulators in SOI by computer simulation. Journal of Lightwave Technology. 2000;18(3):443.

[14] Cherchi M, Bera A, Kemppinen A, Nissilä J, Tappura K, Caputo M, et al. Supporting quantum technologies with an ultralow-loss silicon photonics platform. Advanced Photonics Nexus. 2023;2(2):024002-2.

[15] Baehr-Jones T, Ding R, Liu Y, Ayazi A, Pinguet T, Harris NC, et al. Ultralow drive voltage silicon traveling-wave modulator. Optics express. 2012;20(11):12014-20.

[16] Thomson D, Zilkie A, Bowers JE, Komljenovic T, Reed GT, Vivien L, et al. Roadmap on silicon photonics. Journal of Optics. 2016;18(7):073003.

[17] Zhang S, Cui K, Pan X, Li X. Real-time Bias Control Technique of Multistage Mach-Zehnder Modulators with Ultra-High Extinction Ratio Based on Dither Signal. Journal of Lightwave Technology. 2025.

[18] Mardoyan H, Jorge F, Destraz M, Duval B, Bitachon B, Horst Y, et al. Generation and transmission of 160-GBaud QPSK coherent signals using a dual-drive plasmonic-organic hybrid I/Q modulator on silicon photonics. In: Optical Fiber Communication Conference. Optica Publishing Group; 2022. p. Th1J-5.

[19] Zhang Z, Huang B, Wang Q, Chen Z, Li K, Zhang K, et al. Polarization-insensitive silicon intensity modulator with a maximum speed of 224 Gb/s. Photonics Research. 2025;13(2):274-85.

[20] Yue H, Fu J, Zhang H, Xiong B, Pan S, Chu T. Silicon modulator exceeding 110 GHz using tunable time-frequency equalization. Optica. 2025;12(2):203-15.

[21] Yu Z, Tu D, Guan H, Tian L, Jiang L, Li Z. Design of silicon traveling-wave Mach-Zehnder modulators with transparent electrodes. Optics Express. 2025;33(1):1237-48.

[22] Png CE, Sun MJ, Lim ST, Ang TY, Ogawa K. Numerical modeling and analysis for high-efficiency carrier-depletion silicon rib-waveguide phase shifters. IEEE Journal of Selected Topics in Quantum Electronics. 2016;22(6):99-106.

[23] Liu A, Liao L, Rubin D, Nguyen H, Ciftcioglu B, Chetrit Y, et al. High-speed optical modulation based on carrier depletion in a silicon waveguide. Optics express. 2007;15(2):660-8.

[24] Thomson DJ, Gardes FY, Fedeli JM, Zlatanovic S, Hu Y, Kuo BPP, et al. 50-Gb/s silicon optical modulator. IEEE Photonics Technology Letters. 2011;24(4):234-6.

[25] Soref R, Bennett B. Electrooptical effects in silicon. IEEE journal of quantum electronics. 1987;23(1):123-9.

[26] Png CE, Chan SP, Lim ST, Reed GT. Optical phase modulators for MHz and GHz modulation in silicon-on-insulator (SOI). Journal of lightwave technology. 2004;22(6):1573.

[27] Reed GT, Mashanovich G, Gardes FY, Thomson D. Silicon optical modulators. Nature photonics. 2010;4(8):518-26.

[28] Azadeh SS, Nojić J, Moscoso-Mártir A, Merget F, Witzens J. Power-efficient lumped-element meandered silicon Mach-Zehnder modulators. In: Silicon Photonics XV. vol. 11285. SPIE; 2020. p. 65-75.

[29] Xu H, Li X, Xiao X, Li Z, Yu Y, Yu J. Demonstration and characterization of high-speed silicon depletion-mode Mach–Zehnder modulators. IEEE Journal of Selected Topics in Quantum Electronics. 2013;20(4):23-32.

[30] Liu Y, Li H, Li Y, Li H, Hao Y, Qin L, et al. High-speed electro-optic modulator with group velocity matching on silicon substrate. Frontiers in Bioengineering and Biotechnology. 2025;13:1626017.

[31] Hu H, Liu S, Ma X, Dong R, Chen H, Fang Q. Large wavelength bandwidth Mach-Zehnder modulator based on optical intensity equalization structure. Frontiers in Physics. 2022;10:1017794.

[32] Rahim A, Hermans A, Wohlfeil B, Petousi D, Kuyken B, Van Thourhout D, et al. Taking silicon photonics modulators to a higher performance level: state-of-the-art and a review of new technologies. Advanced Photonics. 2021;3(2):024003-3.